# Search engine effects on news consumption: ranking and representativeness outweigh familiarity in news selection

Ulloa, Roberto[1]     Kacperski, Celina Sylwia[2,3]

[1]GESIS – Leibniz Institute for the Social Sciences, Computational Social Science, Unter Sachsenhausen 6-8, 50667 Köln, Germany

[2]University of Mannheim

[3]Seeburg Castle University

## Abstract

Online platforms have transformed the way in which individuals access and interact with news, with a high degree of trust particularly placed in search engine results. We use web tracked behavioral data across a 2-month period and analyze three competing factors, two algorithmic (ranking and representativeness) and one psychological (familiarity) that could influence the selection of news articles that appear in search results. Participants' (n=280) news engagement is our proxy for familiarity, and we investigate news articles presented on Google search pages (n=1221). Our results demonstrate the steering power of the algorithmic factors on news consumption as compared to familiarity. But despite the strong effect of ranking, we find that it plays a lesser role for news articles compared to non-news. We confirm that Google Search drives individuals to unfamiliar sources and find that it increases the diversity of the political audience to news sources. With our methodology, we take a step in tackling the challenges of testing social science theories in digital contexts shaped by algorithms.

## Main

Online platforms have expanded the ways in which individuals access and interact with news (Möller et al., 2020). Across a variety of countries, websites have become the most important source for news (75%); about 25% of news users name search engines as their main way of coming across news online, with 25% direct and another 26% via social media (Newman et al., 2021). There is a need to understand how individuals seek information and how these decisions affect their behavior (Loecherbach et al., 2021; Sharot & Sunstein, 2020). Search engines are of particular interest: with their role as gatekeepers (Nechushtai & Lewis, 2019), they are in the privileged position of fulfilling an individual's need for information by preselecting, sorting, and presenting resources from the millions available on the Internet.

The trust that users place in search engines is evidenced by how often they select the top items of the results provided by the engine (Pan et al., 2007; Urman & Makhortykh, 2021). The ranking effect is so widely accepted that heavy weights are used in studies that investigate biases in search engines (Kulshrestha et al., 2017; Robertson et al., 2018). In general, individuals use search results as a measure for content quality and relevance (Edelman, 2021; Keane et al., 2008; Schultheiß et al., 2018; Urman & Makhortykh, 2021), influencing judgments, decisions and behavior related to topics ranging from purchases (Ghose et al., 2014) and writing (Novin & Meyers, 2017) to correctness of health information (Kammerer & Gerjets, 2012; Lau & Coiera, 2009) and voting preferences (Epstein et al., 2017; Epstein & Robertson, 2015; Zweig, 2017).

Despite the strong effect of ranking, it is important to not overlook other potential factors that might alter the selection of results. Some studies indicate that using search engines is associated with more diversity of news sources visited by individuals (Fletcher et al., 2021; Fletcher & Nielsen, 2018; Scharkow et al., 2020), while many have analyzed representativeness of sources in search results (Diakopoulos & Koliska, 2017; Haim et al., 2018; Jiang, 2014; Nechushtai & Lewis, 2019; Puschmann, 2019; Steiner et al., 2020; Urman et al., 2021). Yet, to the best of our knowledge, no one has yet investigated how representativeness of news sources affects individuals' selection of news articles. This is an important consideration because the probability with which a user selects a domain might be affected by the frequency with which it is displayed on the search result page.

Most interestingly, personal preferences should influence the result selection, but they remain understudied due to technical difficulties of collecting behavioural data that presents a comprehensive picture of what individuals see and do online. One prominent recent exception used web tracking in order to investigate individuals' preferences towards politically congruent news sources: they found that individuals in the US who strongly identified as Republican engaged with more partisan and unreliable news than Democrats did (Robertson et al., 2021).

Psychological theory can be enlisted to explain some findings in the online news consumption literature. The strong effect of ranking on choice is in line with findings on default effects, well-supported across a variety of behavioral experimental studies (Hummel & Maedche, 2019; Jachimowicz et al., 2019). Users have been shown to generally approach political messages with a confirmation bias (Knobloch-Westerwick et al., 2015), in line with findings of political congruency in their online news selection (Robertson et al., 2021). More specifically related to a hypothesis of familiarity predicting online news selection, a mere-exposure effect would explain a preference of individuals for news sources that they are already familiar with (Montoya et al., 2017), similar to the function of the familiarity heuristic, which in consumer behavior explains why individuals are more likely to consume products of the same brand (Park & Lessig, 1981).

In the context of the argument made for theory development in environments shaped by algorithms (Wagner et al., 2021), neither algorithmically decided factors nor psychological phenomena such as familiarity should be explored in isolation. Thus, supported by the above theories and empirical data, we study three factors using linear mixed-effect models (controlled for surveyed individual characteristics) to investigate how individuals select news articles from a list of search results: two algorithmic (ranking and source representativeness) and one psychological (familiarity), and study their potential interactions. In a novel manner, we explore web tracking data in the form of a behavioral field study to analyze the relation between these three factors. Access to the full HTML of individuals' visited pages and tab activity of the browser allowed us to examine not only their selections, but also the alternative options to which participants were exposed.

To investigate familiarity, i.e., the participants' acquaintance with the news sources they are presented in the search results, we propose engagement as a proxy. We measure it using a section of the browsing history that is independent from the analyzed news articles that are selected (and visited) in the search results. This allows us to quantify the existent relationship between the individual and a given news source through the number of visits of an individual to a given news domain; thus, capturing three modes of news engagement: routinary visits, social media referrals, and intentional search (Möller et al., 2020). Accessing the browsing history provides a metric that is more robust as it is a behavioral measure as opposed to self-reports, including indirect proxies such as political alignment.

Our results do not support the presence of a familiarity effect in the selection of news articles in Google search results. Instead, we corroborate previous evidence that Google ranking has a large effect on news consumption (Pan et al., 2007; Schultheiß et al., 2018; Urman et al., 2021). We also observe a small backlash effect of result representativeness, i.e., more represented sources are less likely to be clicked on.

Additionally, we show that ranking matters less for news articles compared to non-news selection, suggesting that individuals are more careful when selecting news. We also demonstrate that traffic driven by Google Search increases the diversity of sources that individuals engaged with, as well as the diversity of the partisanship audience that visited news sources. Our method for identifying Google Search referrals is novel and precise; not only do we exploit the traceability of the navigation through the tab activity of the browser, but we also verify that the visited URLs exist in the search results of the HTML.

In the last decade, the academic debate about online platforms and their propensity to create or maintain filter bubbles has been vehement (Flaxman et al., 2016; Fletcher et al., 2021; Haim et al., 2018; Keijzer & Mäs, 2022; Nechushtai & Lewis, 2019; Pariser, 2011; Puschmann, 2019; Robertson et al., 2018; Stier et al., 2022). In line with recent empirical evidence (Fletcher et al., 2021; Haim et al., 2018; Nechushtai & Lewis, 2019; Puschmann, 2019; Robertson et al., 2021; Stier et al., 2022), we find no support in favor of filter bubbles. We confirm that individuals place their trust in Google, and that in turn, Google steers them towards sources that are different to their routinary visits. In this process, we tackle the challenge of developing reliable measurement models when integrating social science theory with digital behavioral data (Wagner et al., 2021).

## Results

**Familiarity with news sources does not predict news article selection, while rank and representativeness do**

We investigated which of three relevant factors best predict news article selection in Google results: *news* (source) *engagement*, *ranking*, or *representativeness*. We use *news engagement* as a proxy for a participant's familiarity with each news source. We define *news engagement* as a proportion[1], the number of visits of an individual to a news source divided by their total number of news visits. The metric for *ranking* refers to the position of a news article within a search page. The metric for *representativeness* is a proportion, the frequency of occurrence of a domain within a search page divided by the number of results. The dependent variable is binary (0,1), representing news articles that appear in the first Google Search results page: we coded each available news result by indicating whether the participant selected it (1) or not (0).

---

[1] The authors are aware of the potential skewness issues of proportions in cases when news visits are low; we therefore additionally modelled absolute visits controlled by the total number of visits. This did not substantially change the results reported in this section.

The web browsing data is split in half to resolve time-dependence between news engagement and selection. For news engagement, we only consider data before the cut-off date (Sept 19th, 2021, 15:15:00, see [News articles results and news domain visits data sets](#)). For Google Search selection, we only consider data after this cut-off date, which comprises 2754 results across 611 search pages with an average of M = 4.51 results per page (Mdn = 4.0, SD = 1.97).

We find a significant effect of *ranking* (OR = 2.95, CI = 2.61, 3.34], $p < .001$), and *representativeness* on article selection (OR = .77, CI = [.66, .90], $p < .001$) (see [Modeling news article selection](#) for details on the fitted models, and Supplementary Material S1 for a full regression summary). We do not find a significant effect of *news engagement* on article selection (OR = .93, CI = [0.81, 1.08], $p = .351$).

We then conducted a variety of tests to check the robustness of our effects. Firstly, we calculated a *Google independent news engagement* measure, i.e., we excluded all traffic referred by Google when calculating news engagement, and a *Google associated news engagement*, i.e., we included only traffic referred by Google. Then, in the linear model, we replaced *news engagement* with either *Google independent news engagement*, or *Google associated news engagement*. In both replacements, we corroborated the above results (see Supplementary Material S2 and S3).

Secondly, we tested whether a lack of power might be the reason for the non-significance of the effect of *news engagement* by using a larger data set. For this extended model, we included all data (ignoring the split by cut-off date), comprising in total 5380 Google results (news articles) across 1221 search pages, with an average of M = 4.41 results per page (Mdn = 4.0, SD = 1.97). With this larger data set, we replicated the same pattern as in the previous model: significant effects for *ranking* (OR = 3.1, CI = [2.83, 3.39], $p < .001$) and *representativeness* (OR = .75, CI = [0.65, 0.85], $p < .001$), and, additionally, a significant interaction of them (OR = 1.15, CI = [1.01, 1.32], $p = .038$). We again did not find a significant effect of *news engagement* on article selection (OR =1.03, CI = [0.95, 1.12], $p = .468$) (see Supplementary Material S4-S6).

**Google Search increases source diversity in news consumption**

*Ranking* and *representativeness* are both determined by Google's algorithms – so the results above demonstrate the leverage that Google Search has in driving users towards news sources. We investigated the implications of this leverage by looking more closely at the source diversity of news visits across the entire tracking period.

We logged 75911 total visits to news domains. Of these, 13018 were referred directly or indirectly by Google services (see Methods), including 1269[2] news articles specifically by Google Search. 62893 news pages were visited by participants independently of Google.

We first investigated individuals' news consumption assuming an internet without Google. Plot A in Figure 1 presents the distribution. The visible skew indicates that participants repeatedly visited their own preferred domains: on average, each participant's top-3 news domains concentrate 87.90% (Mdn = 95.65, SD=15.95) of their news consumption.

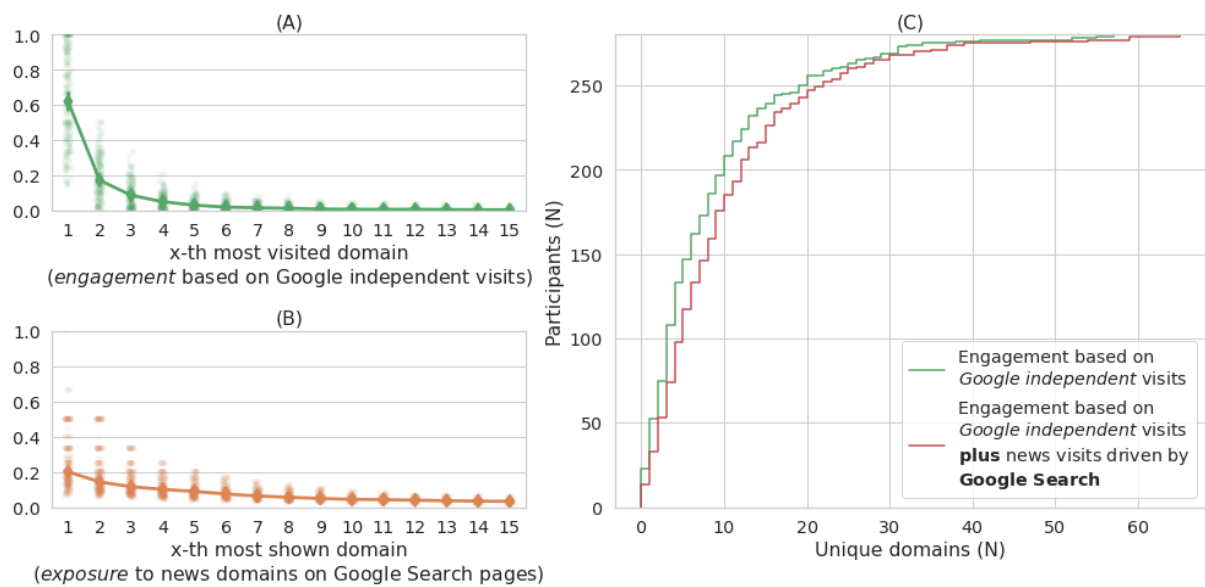

**Figure 1. News diversity with and without Google Search.** The plots display distributions of (A) rank of most visited domain, (B) rank of most shown domain on the search results, and (C) number of unique domains visited. Plot A shows the proportion of domain visits to total visits (Y-axis) by most visited domains (X-axis, truncated to top-15 out of 57); green data points represent the participants, and the green line is the median. Plot B shows the likelihood of a domain to be shown on the first page of Google Search results (Y-axis) by most shown domains (X-axis, truncated to top-15 out of 63); red data points represent the participants, and the red line is the median. Vertical lines in plot A and B display confidence intervals at 95%. Plot C shows the cumulative distribution of participants (Y-axis) to the number of unique domains (X-axis): the green line represents Google independent visits, red includes visits driven by Google Search.

---

[2] The number of search pages (N=1221) is different from the selected news articles (N=1269) because participants could have selected more than one article on a given search page. The referrals to news articles reported included here only include those in dataset A (see Methods section).

We then investigated how Google Search exposes users to news. Plot B presents the distribution of how often news domains are presented in the search results. The distribution is flatter (compared to Plot A); on average, the top-3 most displayed news domains on each participants' Google Search page concentrate 56.09% (Mdn = 47.14, SD=27.11) of their news exposure.

Finally, we tested how Google Search contributes to source diversity of news consumption. Plot C presents the increase in news visits when adding Google Search to participants' otherwise Google-independent browsing experience. We find a statistically significant difference between the two, $D_{KS}(278) = .125$, $p < .0251$, with an average increase of ~32.35%, corresponding to 2.88 (Mdn = 2, SD = 2.94) added unique domain visits (increased from 8.89, Mdn = 6, SD = 9.25). This increase in source diversity is achieved with only 2.02% additional visits (1269 to the 62924 Google independent visits).

**Google Search increases audience diversity in terms of partisanship**

We explored the political alignment of participants in relation to their news consumption to better understand how Google Search shapes the audiences of different news sources. We calculated a measure of audience diversity using participants' political alignment (scale from 0 to 10). A higher variance here indicates that participants from both the left and right end of the spectrum visited a news source.

Again, we first looked at browsing behavior supposing an internet environment without Google. Audience diversity had a mean of 2.92 (SD= 1.81), indicating a relatively homogenous readership to each news source. We then looked at browsing behavior through Google Search, which had a mean of M = 3.59 (SD = 2.84), significantly higher, $t(236) = 2.22$, *two-sided p* = .027. This indicates that Google Search drove a more diverse partisanship audience to news sources. This difference is even larger if we look at the page views (visits) instead of unique participants; $t(271) = 4.49$, *two-sided p* < .0001. Figure 2 illustrates this result.

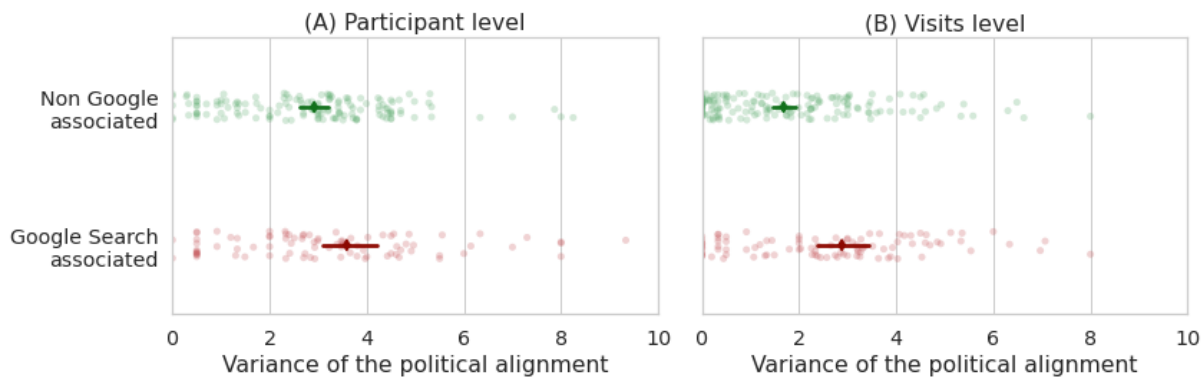

**Figure 2. Partisanship audience of Google independent and Google Search visits.** The left plot (A) shows the variance of the political alignment (X-axis) of the unique participants that visited each news domain according to the association of the visit to Google (Y-axis): Google independent and Google Search associated. The right plot (B) shows the equivalent plot for each visit (page view). Confidence intervals at 95% are represented.

**Comparing news and non-news: ranking matters less for news articles**

Consistent with previous literature (Wojcieszak et al., 2021), we found that news consumption differs widely across participants. Across the tracking period, the median frequency of news visits in our subsample was 90.5 (M=271.24, SD=448.92, N=280); this represents 2.32% (Mdn=1.02, SD=3.89, N=280) of the overall browsing. 10.36% (29 of 280) of participants had less than 10 news visits and 6.76% (19 of 280) of participants visited more than 1000 news sites in the ~2 months period. Over this period, 8.21% (23 of 280) did not visit any news sites by means other than Google (i.e., without being referred by Google).

We investigated how Google Search ranking differs in its effect on participant's choices in news articles versus non news articles. We first looked at the top position (1st ranking). We found that non-news are selected in 46.71% of cases (Mdn = 45.61, SD = 16.32), while news are selected 31.96% of the time (Mdn = 25, SD = .33), a difference that is statistically significant t(557) = 6.68, *two-sided p* < .0001. Plot (A) in Figure 3 shows the proportions for each Google Search ranked result (black and red lines). We hypothesized that this effect might be driven by Google ranking news articles lower than non-news on average. Contrary to our hypothesis, we find that the top results are slightly more likely to showcase news, $r_s$(1928) = -.11, *two-sided p* < .0001[3].

To get a better idea of individuals' browsing behavior overall, we tested whether participants' frequency of news visits correlated with their non-news visits; we found that they correlated weakly, $r_s$(278)=.34, p <0.001 (plot B, Figure 3). This indicates that participants that consume more news are not necessarily those that browse the most. For example, in plot B, we can see that many participants with few news visits have as many other site visits as those with over 1000 news visits.

---

[3] This test includes only Google Search pages with exactly 10 results to avoid over-representation of top-results due to incomplete pages.

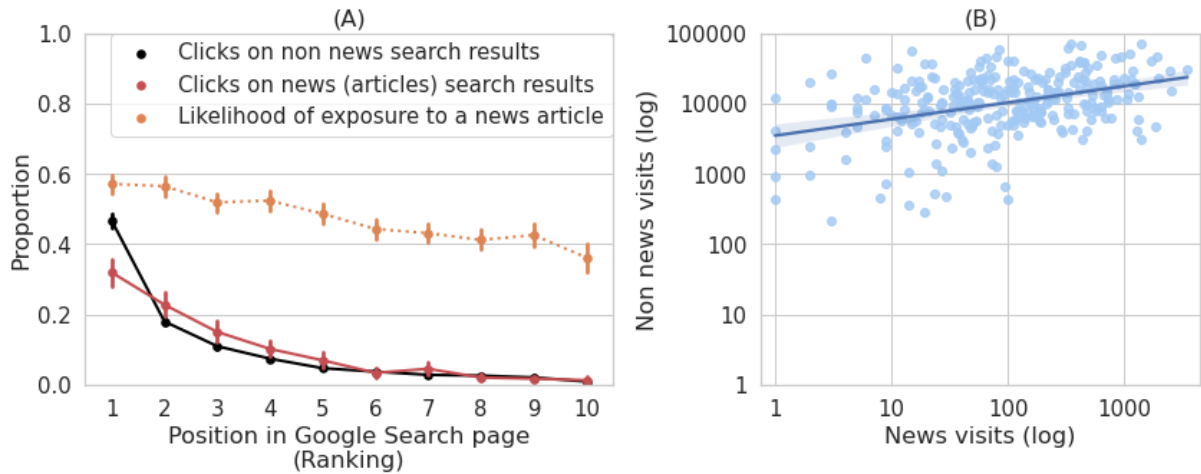

**Figure 3. Left: selection of results according to Google Ranking (A). Right: relation between news and non-news visits (B).** In plot A, the X-axis indicates Google ranking of results presented on the search page. The Y-axis displays either the proportion of result selections across participants (black and red lines), or the likelihood of being exposed to a result leading to news articles in pages that contain at least one news article (orange dotted line). Vertical lines represent confidence intervals at 95%. The scatter of plot B shows the relationship of participants' number of new visits (X-axis) to the number of non-news visits (Y-axis).

**Familiarity: a matter of consistency and reliability**

In this section, we will further explore the underlying patterns behind our proxy for familiarity: news engagement. In the first results section, we did not find significant evidence supporting the hypothesis that news engagement impacts news selection when individuals are using Google Search. In this section, we show that, comparing behavior across two time periods, participants were not very consistent regarding the news sources they consumed. For the time periods, we employed the same cut-off date as for our first model testing news selection.

First, we found a moderate correlation of the visits for each participant-domain[4] pair, $r_s(62438) = .44$, $p = .0001$. Knowing the leverage that Google has in driving users to news domains, we explored whether the consistency of news consumption is shaped differently depending whether Google services (Google Search, Google Maps, Google News) do or do not drive participants to news. When looked at individuals' news engagement supposing an internet experience without Google (Google independent news), removing all Google referred browsing, we found a stronger correlation; $r_s(62438) = .52$, $p < .0001$.

---

[4] In this dataset, we only included news domains with at least one visit.

These correlations are difficult to interpret because most domains were never visited by most of the participants (i.e., a large amount of 0s in the vectors might inflate the correlations). Thus, we used the Jaccard index (JI), a measure of similarity between datasets to establish how similar participants' news visits were for the two timespans. We defined *thresholds t* and for a given *t*, we only included participants that had at least t visits before and t visits after the cut-off date. We found that participants' behavior was robust (JI > .42, green bars in Plot A of Figure 4), especially considering that news visits are rare for some participants. We also calculated the corresponding RBO (Rank Biased Overlap) (Webber et al., 2010), a similarity metric which uses a parameter p to add a weight of the rank of each domain based on the participants' number of visits. The results for various values of p and t are shown in the green trace in Plot B of Figure 4. The similarity is higher when the ranking is considered, especially for the top positions (lower values of p).

We finally turned our attention to participants' Google associated news visits, i.e. news visits directly or indirectly referred by a Google service; in this case we found a much lower correlation, $r_s(62438) = .29$, $p < .0001$. Consequently, the similarity of participants' news visits was lower for the two timespans: JI < .24 (purple bars in Plot A of Figure 4) and RBO < 0.4 (purple in Plot B of Figure 4). Here, the RBOs were stable for varying p parameter values, indicating that the consistency did not improve even for the most visited domains.

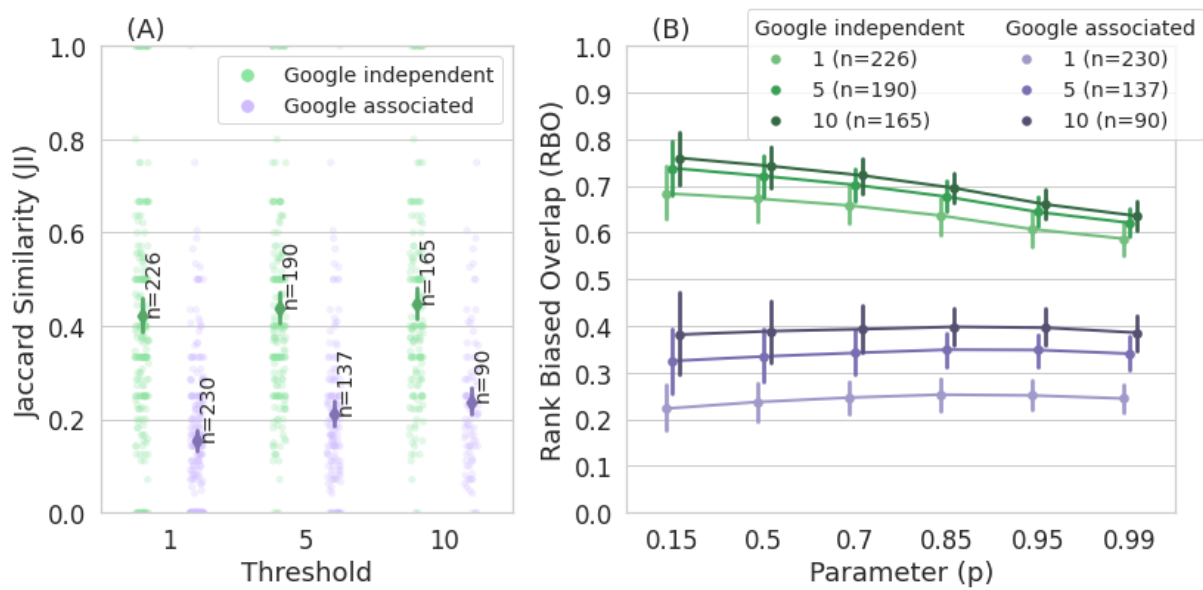

**Figure 4. Similarity of the engagement measure across time.** Plots A and B measure the average similarity of news engagement between visits before and after the cut-off date (calculated per participant). Plot A shows the Jaccard Similarity (JI, Y-axis) of the news engagement for *Google independent* news visits (green), and for *Google dependent* visits (purple); it also shows the JI

similarity of the news domains that were shown in Google results (exposure, orange). The X-axis indicates different thresholds for filtering out participants based on their total visits (the number of participants that fulfil the threshold, n, is displayed over the bar). Plot B shows the corresponding RBO scores (Y-axis) with results for different values of the parameter p (X-axis). The legend for B includes the corresponding thresholds and participant numbers (n). Vertical lines represent confidence intervals at 95%.

The volume of news consumption of the participants was consistent. For the Google independent visits, we ran a Kolmogorov–Smirnov test, and we did not find evidence supporting different distributions between the visits per participants before and after the cut-off date; $D_{KS}(278) = .043$, $p < .960$ (see distribution, Plot A, in Figure 5), and the correlation was very high, $r_s(278) = .85$, $p < .0001$ (see scatterplot, Plot C, in Figure 5). Similar results hold for Google associated visits: we did not find evidence supporting different distributions, $D_{KS}(278) = .043$, $p < .960$ (see Plot B in Figure 5) as well as a high correlation, $r_s(278) = .56$, $p < .0001$ (plot D in Figure 5).

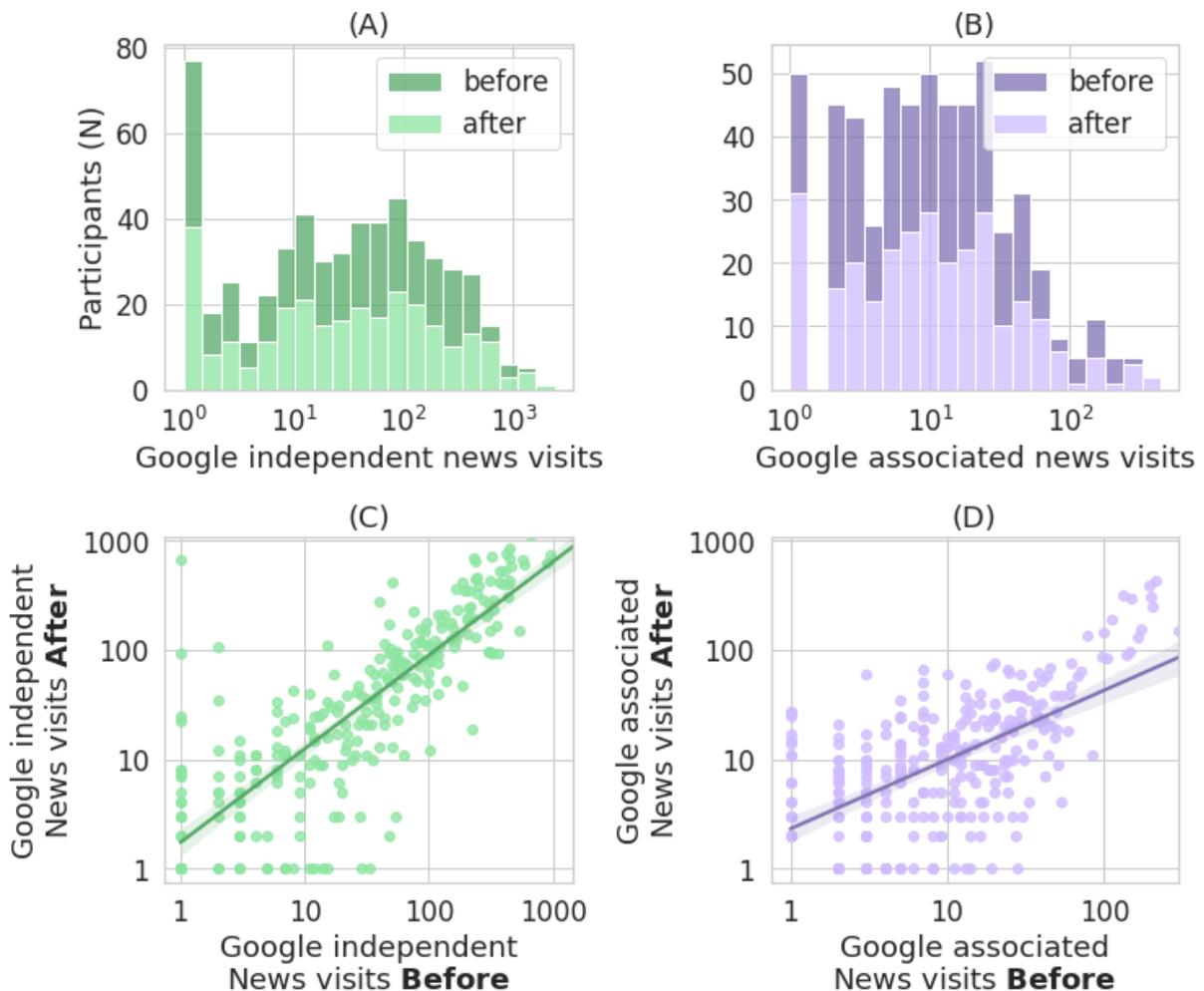

**Figure 5. News related web activity.** Plot A and B show the distribution of participants (Y-axis) according to their number of *Google independent (A) and Google dependent (B)* domain visits (X-axis, log-scale); the dark and light sections (stacked) distinguish between the visits before and after the cut-off date. Plot C and D shows the scatterplots of the visits before and after for Google

independent news visits (C) and Google associated news visits (D), both axes are logarithmic. The semi-transparent bands are confidence intervals at 95%.

The lack of consistency between news domain visits within the Google ecosystem is in line with results we present in the previous sections, which have suggested that Google Search facilitates more diverse news selection for its users. The overall results suggest that while participants' overall volume of news consumption is consistent, and that participants do, on their own, have preferences for familiar news sources (as indicated by the similarity of sources visited across time), this familiarity loses relevance when they browse for news while integrated in the Google ecosystem, and in particularly when choosing their news in Google Search.

## Discussion

We tested the hypothesis of news engagement (as a proxy for familiarity with news sources) as a predictor of news article selection in the Google search engine. We did not find evidence supporting this. Instead, we found a significant effect for two factors that are decided by the search engine alone: the position in which the result is presented (ranking) and the number of times the news source appears (representativeness). While ranking has been demonstrated to play a strong role, we show that its effect is weaker for news article selection compared to non-news selection. This may well suggest a more careful decision-making process of individuals when selecting news (e.g., reading the titles and excerpts more attentively). An investigation of this phenomenon in future research could employ for example analysis of interaction durations or even eye-tracking methodologies.

Surprisingly, representativeness reduced the likelihood of news article selection. This might be an indication that once individuals have decided not to visit a result belonging to a specific news source, they also discard subsequent results from the same source. Usage of this metric could in future research serve as a method to study news avoidance in concrete contexts (defined by query terms) where the individual is proactively seeking information.

In line with previous research, we found that Google Search increases the diversity of participants' news consumption (Fletcher et al., 2021; Fletcher & Nielsen, 2018; Scharkow et al., 2020). It is possible that participants use Google Search when they are actively looking for novel news sources, though we also show that Google Search facilitates a discovery process by presenting a variety of news sources among the results. Additionally, we show that Google Search increases political audience diversity that news sources receive. Given that Google has its own news quality controls in place (Google Developers, 2021), our findings can explain recent research showcasing that political audience diversity can be used as a sign of news source reliability, and that it should be incorporated into ranking algorithms (Bhadani et al., 2022). In our case, it seems it is Google Search who drives this effect. More broadly, researchers should consider that online news browsing behavior is heavily shaped by online platforms. We demonstrated that there are differences in the consistency of our familiarity metric, depending on whether it is measured including traffic referred by Google or not.

We would like to point out several limitations. First, our engagement metric does not fully capture the broader scope of familiarity. It is possible that individuals are familiar and trust a given source, yet they do not regularly consume news. Our measurement does not include non-online activity. And the engagement for the time independent analysis was calculated using an arguably short period of time (~1 month). We hope future data collections will enable analyses across timespans of multiple months up to years, while we highlight that our results did not change when considering a time dependent analysis (~2 months). Second, our analysis is limited to Google, which we chose due to its market dominance (Statcounter, 2021) and findings should not be generalized to other search engines. Third, characteristics of our sample should be kept in mind when interpreting our results. The sample size used in the above research is relatively small. Out of the 739 individuals that participated in the web tracking study overall, only 280 are represented in our subsample. This is due to a relatively low number of news visits (including the ones driven by Google search) in web tracking data - consistent with previous literature (Scharkow et al., 2020; Wojcieszak et al., 2021), and the strict data quality constraints for including a search page for the analysis. Our sample is relatively uniform, only including German individuals with a Chrome or Firefox browser installed on their desktop computers; while this reduces noise, it also affects generalizability of the findings. Finally, many individuals refuse to participate in web tracking studies due to privacy concerns (Makhortykh et al., 2021), which might indicate that the sample is pre-selected based on factors that we don't yet fully understand and cannot control for.

Despite these limitations, the logged web browsing behavior that we capture occurs in a real environment with minimal intervention; we argue that cognitive awareness of the presence of the web tracker is likely to only affect the very initial browsing behavior. Moreover, web tracking studies that include the website content remain rare, and our method is exceptional as it deterministically identifies referrals by tracing the tab activity of the browser and matching the presence of the URLs among the results.

In conclusion, although we find consistent news engagement behavior (outside of traffic directed by Google) that supports our metric as a measure of news familiarity, and that showcases participants' preferences for news sources, we find no evidence that these preferences affect future news choices when searching for news in a search engine. Instead, our findings demonstrate the steering power that Google Search has for news consumption, and that, in contrast to filter bubbles, Google Search rather drives users to novel news sources, with ranking, as expected, majorly impacting individuals' news engagement decisions. Future research into concepts such as news source avoidance, browsing durations and experimental research studying the impact of interventions on search selection might shed further light on other processes that shape the decisions that individuals make about their news diets.

## Methods

**Data sources**

We used data from a web tracking study that collected browsing activity of individuals in Germany between August 19th, 2021 and October 27th, 2021. Participants in a commercial market research panel were invited to a survey (from which we used demographics variables) and to install a desktop browser plugin (Chrome, Firefox) that collected metadata and the raw HTML of pages visited during the collection period (Aigenseer et al., 2019). Depending on participation rates, participants were compensated with 25 to 75 EUR. In total, ~8.36M page visits of 739 participants (M=11311.07, Mdn=7436, SD=13730.45) were recorded. After data cleaning, 1221 Google Search result pages, comprising 5380 news articles, were included in the analysis; 280 participants were represented with at least one search page in which they selected at least one news article (N=1221, M=4.36, Mdn=3, SD=6.63). Of these, 122 identified as women, 155 as men and one as non-binary. 60 participants were from East Germany and 218 from West Germany. 34 participants held at least elementary level education, 93 had a mid-level education and 151 reported a high education level (high school or above). The sample's mean age was 46.73 (Mdn=48, SD=13.96, min=18, max=74). The sample's mean in terms of political alignment was 5.45 (Mdn=6, SD=1.88, left=1, right=10). The 280 participants conducted ~4.27M visits (M=15261.05, Mdn=11286.50, SD=13239.92), of which 75911 (M=271.11, Mdn=90.5, SD=448.59) were to news domains (i.e., news visits), which we identified using three lists (AllSides, 2022; Robertson et al., 2018; Stier et al., 2020).

**Data protection**

Data collection was approved by an institutional ethics committee and conducted in line with institutional data protection and ethics regulations, including the collection of explicit informed consent with regards to participation in the surveys and in the web tracking procedure. The WebTrack plugin avoided collection of sensitive websites and information (pre-generated deny lists), and participants could at any time temporarily de-activate the tracking by switching on a private mode button within the plugin. Because web tracking data inherently carries the potential for de-anonymization, data access is restricted to researchers directly authorized by the project lead. Fully anonymized datasets can be made available upon request to reproduce the here presented analysis and plots.

**News articles results and news domain visits data sets**

Using the web tracking data, we generated five data sets for our analysis, namely: (A) *search news* (articles) *results,* (B) *all news visits*, (C) *Google associated* (news visits), (D) *Google independent* (news visits) and (E) *non-news* (visits). The *search news results* data set contains news articles results that appeared in Google Search pages. The *all news* data set includes all visits to news domains regardless of the way in which participants arrived at them. The *Google associated* data set includes only those news domain visits that were referred by Google services. The *Google independent* data set includes only those news visits that were not referred by Google services. The *non-news* data set includes all website visits that did not correspond to any news domains. The left side (I) of Figure 6 presents a schematic representation of the dataset that we will further explain.

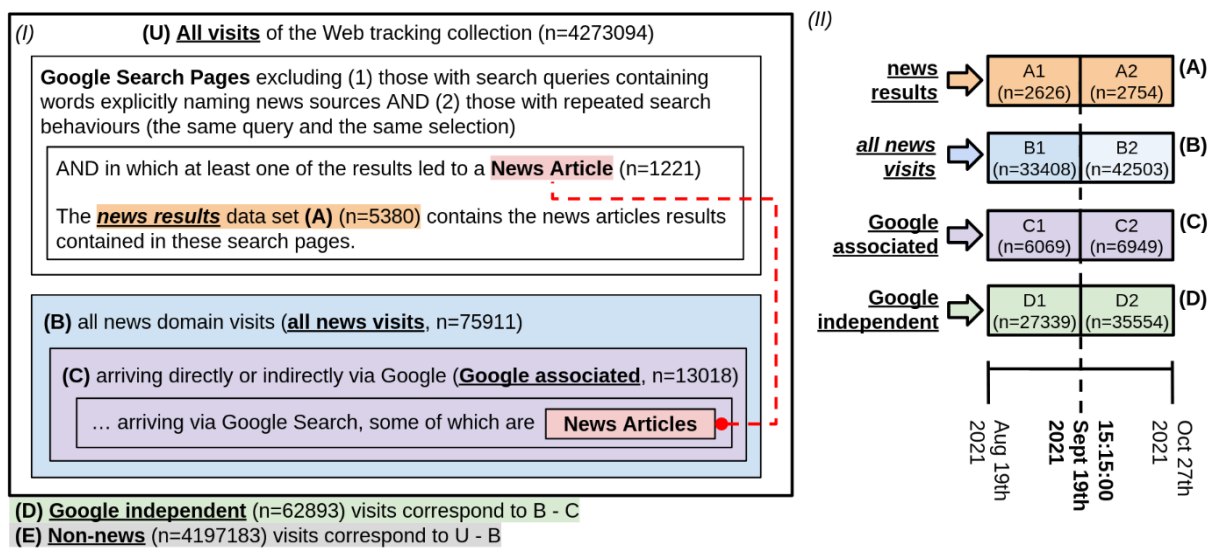

**Figure 6. Diagrams of the data sets extracted from the web tracking data.** The left figure (I) represents all the pages in the web tracking collection. Rectangles inside represent subsets: Google Search Pages related subsets, and news related subsets. The colored rectangles and upper-case letters in parenthesis highlight relevant data sets: blue (B) for *all news* domain visits; purple (C) for the *Google associated* news domain visits; green (D) for the *Google independent* news domain visits; red

(D) for the Google Search pages used to generate the news results data set, and orange (A) for the set of Google Search results that correspond to news articles. The diagram describes basic conditions used to filter the data sets. The right figure (II) represents the way in which the data sets are split, so that it is possible to create time independent models by using the data before the cut-off date (Sept 19th 2021, 15:15:00) to measure predictors (independent variables) that could explain responses (dependent variables) in the data after the cut-off date. Colors in Figure (I) and Figure (II) are aligned and used consistently throughout figures in this paper.

To generate the *search news results* data set (identified with the letter A, orange, Figure 6), we followed these steps:

First, we extracted (from the *all news* subset, explained below) the news pages at which participants arrived directly from Google results. Second, we parsed the HTML of the pages corresponding to the Top-10 Google Search results and extracted related data (including their ranking, and URL). Third, we flagged the results that were selected ("clicked on"); a result was considered selected if the URL of a subsequent news visit corresponded to the URL of the result (redirects and cookie consent pages were omitted to the corresponding sequence between the result selection and the news article). Fourth, we discarded search pages resulting from search query terms that included references to news sources; we automatically discarded queries that included news root and main domains. For example, "elections 2021 bild" or "elections 2021 bild.de" would be discarded, because "bild"/"bild.de", referring to the German newspaper Bild, formed part of the query term. We manually inspected the search queries to include typos (e.g., "bild:de" instead of "bild.de") as well as other news source names (e.g. "Frankfurter", which refers to the German newspaper "Frankfurter Allgemeine" and whose domain is faz.net). Fifth, we manually annotated the URLs of the Google results, and (1) discarded results that did not correspond to news articles, and (2) full Google Search pages that did not contain at least two news articles of different domains (otherwise making it impossible to compare news sources). For the annotation, a URL was considered an article if (1) it was hosted by a news domain, and (2) the content (including text and video) informed potential readers about a non-fictional topic (opinion pieces such as commentaries or documentaries are included). One of the authors inspected all URLs (reviewing the content when it was unclear); the remaining ambiguous cases were discussed and resolved with the other author. Sixth, we removed repeated search behaviors, i.e., a participant searching for the same query term and selecting the same result.

To generate the *all news* data set (identified with the letter B, blue, in Figure 6), we used three lists of news domains (AllSides, 2022; Robertson et al., 2018; Stier et al., 2020) and so identified all news related visits. The *Google associated* data set (identified with the letter C, purple, in Figure 6) is a subset of the *all news* data set in which we only included visits that were driven, directly or indirectly, by Google (including subdomains such as maps.google.com, and country level domains such as google.de or google.co.uk). We defined that a participant arrived directly from page $x$ to page $y$ if the browser tab that displayed x was opened from a tab displaying y; we then pose that this is a direct path from $x$ to $y$ (denoted $x \to y$). We defined that a participant arrived indirectly from page $x$ to page $y$ if (a) there existed a sequence of direct paths that led from $x$ to $y$, i.e., $\exists \{z_1, \ldots, z_n\} : x \to z_1, z_1 \to z_2, \ldots, z_{n-1} \to z_n, z_n \to y$, and (b) $z_j \in \{z_1, \ldots, z_n\}$ belonged to the same web domain $d$. The path $x \to y$ exists if $y$ was opened from a tab displaying $x$. The reason for this is that the browser extension can identify the tab from which a visit was initiated (whether if the visit is open in the current tab or in a new one - including tabs on new browser windows), but it cannot identify if the visit was caused by clicking on a link or because the participant manually typed an URL in the navigation bar. For the same reason, we excluded paths originating from a Google home page URL (e.g., https://www.google.com/), as we can safely assume that home pages do not contain news links - it is more likely that a participant's browser is configured to have the Google home page as its default page.

The *Google independent* data set (identified with the letter D, green, in Figure 6) was derived by excluding the rows in the *Google associated* data set from the *all news* data set (i.e., B - C).

We considered statistical tests time-independent by using the portion of the *all news* and *Google associated* data sets that was collected before the cut-off date (Sept 19th, 2021, 15:15:00), and the *news results* data set that was collected after the cut-off date. For example, our main model measures the *news engagement* in the data set B1, and it is used to predict the news article selection in the A2 data set. The cut-off date (Sept 19th, 2021, 15:15:00) splits in half Google Search pages corresponding to the *news results*.

The *non-news* data (identified with the letter E, gray, in Figure 6) is derived by excluding the rows in the *all news* data set from the entire Web tracking collection.

**Definition of news engagement, representativeness, and ranking.**

**News engagement.** We used participants' news visits to construct a behavioral indicator of news source engagement, which measures the strength of the relation between a news source and a participant based on their browsing history. Let $u_1, u_2, \ldots, u_k$ be the list of urls corresponding to the news visits of a participant $p$ - then the engagement of $p$ with the domain $d$ is given by:

$$\frac{1}{k}\sum_{i=1}^{k} S(u_i, d)$$

where $S(u, d) \epsilon \{0,1\}$ indicates if the domain of the url $u$ corresponds to $d$ (1), or not (0).

**Representativeness.** Let $T = t_1, t_2, \ldots, t_k$ be the list of urls corresponding to the results shown in one Google Search result page. The representativeness of a news source, which domain is $d$, is given by:

$$\frac{1}{k}\sum_{i=1}^{k} S(t_i, d)$$

where $S$ was defined previously. Note that $k$ is usually 10, as it corresponds to the results displayed on the first result page.

**Ranking.** The ranking is given by the position of the news article on the Google Search page. Note that, idiomatically, the result with lowest ordinal value (i.e. top-1 result) is referred to as the result with the highest ranking. For the purposes of readability, we keep this convention in all our analyses, including the odds ratios (OR) reported in the binomial linear mixed-effect model regression tables, i.e., positive values indicate that a higher ranking increases the probability of a result being selected.

**Modeling news article selection**

To answer the research question that motivated this investigation, we used a binomial linear mixed-effect model fitting the interaction between the 3 study factors (ranking, representativeness and engagement) while controlling for the following individual characteristics that were centered and scaled before use: is a woman (yes: 1, no: 0), education (low: 0, middle: 1, high: 2), from region (East Germany: 0, West Germany: 1), age (ordinal), political alignment (10 point scale), and election interference (as a major election occurred during data collection): whether the search was performed on the day of the election (yes: 1, no: 0), and the day before or after the election (yes: 1, no: 0). We defined random intercepts for repeated measures (participant, search page, and news domain). The dependent variable was binary, with 1 indicating that the participant selected ("clicked on") the result, and 0 that they did not.

To avoid time dependencies (and thus correlational effects), we split the collection into before and after a cut-off date (Sept 19th, 2021, 15:15:00). We used the data before the cut-off date to calculate the news engagement, and the Google *news results* after the cut-off date (see A2 on the right side (II) of Figure 6). Data set A2 comprises 2754 search results representing 611 search pages, and 194 participants: 90 women, 1 diverse person, 103 men; 41 from East Germany, 152 from West Germany; 23 of low, 65 middle and 105 of high education. The sample's mean age was 46.26 (Mdn=47, SD=13.89, min=18, max=73). The sample's average political alignment was 5.34 (Mdn=6, SD=1.93, left=1, right=10). We fit the model three times for the three data sets according to Figure 6: *all news visits* (B1), *Google associated* (C1) and *Google independent* (D1).

To further explore the main result, we fit a time dependent model using the data of the entire period, i.e., we used all the Google *news results* in data set A and calculated the engagement using *all news visits* data set (B), while excluding visits that were, directly or indirectly, driven by the selected *news results* (A) (i.e. excluding cases that we are inferring in the *news engagement* predictor). The *Methods - Data sources* section already reports relevant demographic distributions for this dataset.

**Post hoc analysis**

**Source diversity of the consumed news.** To study the diversity of the news sources that the participants visited, we looked at the number of unique domains. As a baseline, we first looked at the diversity in the Google independent data set (data set D). We then reported the diversity attributed to Google Search, i.e., we measured the increase of diversity when the domains of the search results selected (i.e., "clicked on") by the participants (and consequently also visited) in the *news results* data set (A) were added to the domain visits of the *Google independent* data set (D). We also measured the diversity of the exposure in Google Search by looking at all the domains in the *news results* (data set A).

**Partisanship audience of news sources.** To explore how Google shapes the audience that news sources receive, we measured the variance of the political alignment of the participants that visited the news domains. Accordingly, we analyzed two types of visits: *Google independent* visits (dataset D), and the specific news visits resulting from the selection among the Google Search results (Google Search driven visits). We used variance because Bhadani et al. (2022) found it to be the best measurement of dispersion to predict the quality of a news source.

**Ranking relevance for news articles.** To analyze the importance of ranking for the results containing news articles, we compared the ranking of the news articles selected by the participants (in the *news results* data set A) with the ranking of non-news results selected by the participants, i.e., including all Google Search pages in the web tracking collection. First, we parsed the HTML of the pages corresponding to the Top-10 Google Search and extracted all their results (including their ranking, and URL). Second, we identified those results that were selected (clicked on), so checked if there exists $x \to y$ where $x$ is the Google Search Page and $y$ the page visited after. Third, we discarded search pages resulting from search query terms that included any domains present in our collection ("xyz wikipedia" would be discarded because "Wikipedia" is included in the term), as we considered this a strong signal of a predisposition towards a certain domain. Fourth, we removed repeated search behaviors, i.e., a participant searching for the same query term and selecting the same result twice. We used a two-sample t-test to establish if there was a significant difference between selecting the first search result for news and non-news, and reported a Spearman' rank correlation $r_s$ to test whether the search engine ranking was biased towards news domains.

**Consistency of news engagement.** We used two measures of similarity to compare two lists of visits: Jaccard similarity (aka Jaccard Index, JI) and Rank Biased Overlap (RBO). Let $F$ and $G$ be, respectively, the sets of news domains corresponding to the lists of visits, then the Jaccard similarity between $F$ and $G$ corresponds to:

$$J(F, G) = \frac{|F \cap G|}{|F \cup G|}$$

By definition, $J(F, G) = 1$ if $F$ and $G$ are both empty. Jaccard only considers the overlap in the domains and ignores the importance (rank) of each domain according to, for example, the frequency of visits. Thus, we complemented the JI using RBO. Let $F$ and $G$ be sequences of domains ordered by their rank (i.e., by the number of times each domain was visited), then

$$RBO(F, G, p) = (1 - p) \sum_{d=1}^{\infty} p^{d-1} \cdot A(F_d, G_d)$$

where $F_d$ and $G_d$ denote the sequence until the element in the position $d$, the parameter $p$ determines the weight given to the top results (the lower, the more importance is given to the top results), and $A$ is the proportion of agreement until the element in the position $d$, formally:

$$A(F_d, G_d) = \frac{|F \cap G|}{d}$$

In all cases, the similarity values were calculated within participants and 95% bootstrapped confidence intervals are presented in the plots.

We used these similarity metrics to assess the consistency across time of the *news engagement* measurement as a predictor by comparing the Google associated visits before the cutting date (data set C1) and after the cutting date (C2), and the Google independent visits before the cutting date (data set D1) and after the cutting date (D2).

**Correlations of visits.** To validate the findings, we used Spearman' rank correlations (denoted by $r_s$) and explored the connection between the within visits in two datasets. Spearman correlation is well-suited to study the relation between the frequency distributions of the domain visits, as the domains are sorted (i.e., ranked) according to their number of visits (i.e., the most visited domain has a higher rank). We report correlations between domain visits within participant (participant-domain pair) in the following datasets: (1) before and after the cut-off date in all news visits (B1 vs B2), (2) in *Google independent* visits (D1 vs D2), (3) and in *Google associated* visits (C1 vs C2). We also report the overall participant visits to news in the following datasets: (1) before and after the cut-off date in *Google independent* visits (D1 vs D2), (2) and in *Google associated* visits (C1 vs C2).

**Comparison of distributions.** We used a Kolmogorov–Smirnov test of equality of distribution (the statistic "difference" is denoted by $D_{KS}$ in the text) (1) to check whether Google Search driven news visits effectively shifted the diversity of Google independent visits and (2) to check that the splits by cut-off dates were balanced between C1 and C2, and D1 and D2.

WORKING PAPER: SEARCH ENGINE EFFECTS ON NEWS CONSUMPTION

## Supplementary Materials

**S1. Full regression model for engagement calculated on the all news dataset before the cut-off date (see B1 dataset in the Data Sources section of the of the Methods)**

Binomial linear mixed-effect model, fitting the interaction between the 3 factors: *news engagement*, *rank* and *representativeness*. We controlled for individual characteristics (age, sex, region, education, political alignment), an election which occurred during data collection, and we defined random intercepts for repeated measures (participant, search page, and news domain).

|  | is_selected_result | | |
|---|---|---|---|
| **Predictors** | **Odds Ratios** | **CI** | **p** |
| (Intercept) | 0.23 | 0.20 – 0.26 | **<0.001** |
| engagement TI B1 | 0.93 | 0.81 – 1.08 | 0.351 |
| inv rank | 2.95 | 2.61 – 3.34 | **<0.001** |
| news representativeness | 0.77 | 0.66 – 0.90 | **0.001** |
| is female | 0.98 | 0.89 – 1.09 | 0.729 |
| is west germany | 0.94 | 0.85 – 1.04 | 0.239 |
| age | 0.95 | 0.86 – 1.06 | 0.398 |
| education | 0.96 | 0.86 – 1.07 | 0.441 |
| political alignment | 1.05 | 0.95 – 1.15 | 0.378 |
| is during elections | 0.95 | 0.85 – 1.06 | 0.363 |
| is after elections | 1.04 | 0.93 – 1.16 | 0.526 |
| engagement TI B1 * inv rank | 1.06 | 0.92 – 1.23 | 0.416 |
| engagement TI B1 * news representativeness | 0.89 | 0.71 – 1.11 | 0.295 |
| inv rank * news representativeness | 1.07 | 0.91 – 1.25 | 0.420 |
| (engagement TI B1 * inv rank) * news representativeness | 1.14 | 0.88 – 1.47 | 0.321 |
| **Random Effects** | | | |
| $\sigma^2$ | | 3.29 | |
| $\tau_{00\ search\_page}$ | | 0.00 | |
| $\tau_{00\ participant}$ | | 0.00 | |
| $\tau_{00\ coredomain}$ | | 0.00 | |
| ICC | | 0.00 | |
| N $_{search\_page}$ | | 605 | |
| N $_{coredomain}$ | | 152 | |
| N $_{participant}$ | | 192 | |
| Observations | | 2730 | |
| Marginal $R^2$ / Conditional $R^2$ | | 0.266 / 0.267 | |

**S2. Full regression model for engagement calculated on the Google associated dataset before the cut-off date (see C1 dataset in the Data Sources section of the of the Methods)**

Binomial linear mixed-effect model, fitting the interaction between the 3 factors: *news engagement*, *rank* and *representativeness*. We controlled for individual characteristics (age, sex, region, education, political alignment), an election which occurred during data collection, and we defined random intercepts for repeated measures (participant, search page, and news domain).

|  | is_selected_result | | |
|---|---|---|---|
| *Predictors* | **Odds Ratios** | **CI** | ***p*** |
| (Intercept) | 0.22 | 0.20 – 0.26 | **<0.001** |
| engagement TI C1 | 0.90 | 0.77 – 1.05 | 0.192 |
| inv rank | 2.97 | 2.62 – 3.37 | **<0.001** |
| news representativeness | 0.76 | 0.66 – 0.89 | **0.001** |
| is female | 0.98 | 0.89 – 1.09 | 0.752 |
| is west germany | 0.94 | 0.86 – 1.04 | 0.260 |
| age | 0.95 | 0.85 – 1.06 | 0.384 |
| education | 0.95 | 0.85 – 1.06 | 0.394 |
| political alignment | 1.05 | 0.95 – 1.16 | 0.343 |
| is during elections | 0.95 | 0.84 – 1.06 | 0.353 |
| is after elections | 1.04 | 0.93 – 1.16 | 0.500 |
| engagement TI C1 * inv rank | 1.09 | 0.92 – 1.28 | 0.307 |
| engagement TI C1 * news representativeness | 1.08 | 0.94 – 1.24 | 0.253 |
| inv rank * news representativeness | 1.06 | 0.91 – 1.25 | 0.452 |
| (engagement TI C1 * inv rank) * news representativeness | 1.06 | 0.87 – 1.30 | 0.546 |
| **Random Effects** | | | |
| $\sigma^2$ | 3.29 | | |
| $\tau_{00\ search\_page}$ | 0.00 | | |
| $\tau_{00\ participant}$ | 0.00 | | |
| $\tau_{00\ coredomain}$ | 0.00 | | |
| N $_{search\_page}$ | 605 | | |
| N $_{coredomain}$ | 152 | | |
| N $_{participant}$ | 192 | | |
| Observations | 2730 | | |
| Marginal $R^2$ / Conditional $R^2$ | 0.267 / NA | | |

**S3. Full regression model for engagement calculated on the Google independant dataset before the cut-off date (see D1 dataset in the Data Sources section of the of the Methods)**

Binomial linear mixed-effect model, fitting the interaction between the 3 factors: *news engagement*, *rank* and *representativeness*. We controlled for individual characteristics (age, sex, region, education, political alignment), an election which occurred during data collection, and we defined random intercepts for repeated measures (participant, search page, and news domain).

| | is_selected_result | | |
|---|---|---|---|
| *Predictors* | **Odds Ratios** | **CI** | ***p*** |
| (Intercept) | 0.23 | 0.20 – 0.26 | **<0.001** |
| engagement TI D1 | 0.96 | 0.83 – 1.10 | 0.546 |
| inv rank | 2.95 | 2.61 – 3.34 | **<0.001** |
| news representativeness | 0.77 | 0.66 – 0.90 | **0.001** |
| is female | 0.99 | 0.89 – 1.09 | 0.772 |
| is west germany | 0.94 | 0.85 – 1.04 | 0.236 |
| age | 0.95 | 0.85 – 1.06 | 0.387 |
| education | 0.96 | 0.86 – 1.07 | 0.441 |
| political alignment | 1.05 | 0.95 – 1.15 | 0.374 |
| is during elections | 0.95 | 0.85 – 1.06 | 0.368 |
| is after elections | 1.04 | 0.93 – 1.16 | 0.540 |
| engagement TI D1 * inv rank | 1.04 | 0.90 – 1.21 | 0.561 |
| engagement TI D1 * news representativeness | 0.90 | 0.72 – 1.13 | 0.361 |
| inv rank * news representativeness | 1.07 | 0.91 – 1.25 | 0.416 |
| (engagement TI D1 * inv rank) * news representativeness | 1.08 | 0.87 – 1.34 | 0.494 |
| **Random Effects** | | | |
| $\sigma^2$ | | 3.29 | |
| $\tau_{00\ search\_page}$ | | 0.00 | |
| $\tau_{00\ participant}$ | | 0.00 | |
| $\tau_{00\ coredomain}$ | | 0.01 | |
| ICC | | 0.00 | |
| N $_{search\_page}$ | | 605 | |
| N $_{coredomain}$ | | 152 | |
| N $_{participant}$ | | 192 | |
| Observations | | 2730 | |
| Marginal $R^2$ / Conditional $R^2$ | | 0.262 / 0.263 | |

WORKING PAPER: SEARCH ENGINE EFFECTS ON NEWS CONSUMPTION

**S4. Full regression model for engagement calculated on the all news dataset (see B dataset in the Data Sources section of the of the Methods)**

Binomial linear mixed-effect model, fitting the interaction between the 3 factors: *news engagement*, *rank* and *representativeness*. We controlled for individual characteristics (age, sex, region, education, political alignment), an election which occurred during data collection, and we defined random intercepts for repeated measures (participant, search page, and news domain).

| Predictors | Odds Ratios | CI | p |
|---|---|---|---|
| (Intercept) | 0.23 | 0.21 – 0.25 | **<0.001** |
| engagement TD B | 1.03 | 0.95 – 1.12 | 0.468 |
| inv rank | 3.10 | 2.83 – 3.39 | **<0.001** |
| news representativeness | 0.75 | 0.65 – 0.85 | **<0.001** |
| is female | 0.99 | 0.93 – 1.07 | 0.862 |
| is west germany | 0.95 | 0.88 – 1.02 | 0.152 |
| age | 1.01 | 0.93 – 1.09 | 0.849 |
| education | 0.95 | 0.88 – 1.02 | 0.154 |
| political alignment | 1.04 | 0.97 – 1.12 | 0.257 |
| is during elections | 0.95 | 0.88 – 1.02 | 0.136 |
| is after elections | 1.00 | 0.93 – 1.08 | 0.938 |
| engagement TD B * inv rank | 0.99 | 0.91 – 1.08 | 0.837 |
| engagement TD B * news representativeness | 1.01 | 0.92 – 1.11 | 0.825 |
| inv rank * news representativeness | 1.15 | 1.01 – 1.32 | **0.038** |
| (engagement TD B * inv rank) * news representativeness | 0.94 | 0.85 – 1.04 | 0.208 |

**Random Effects**

| | |
|---|---|
| $\sigma^2$ | 3.29 |
| $\tau_{00}$ search_page | 0.00 |
| $\tau_{00}$ participant | 0.00 |
| $\tau_{00}$ coredomain | 0.01 |
| $N$ search_page | 1205 |
| $N$ coredomain | 178 |
| $N$ participant | 275 |
| Observations | 5323 |
| Marginal $R^2$ / Conditional $R^2$ | 0.281 / NA |

- 31 -

**S5. Full regression model for engagement calculated on the Google associated dataset (see C dataset in the Data Sources section of the of the Methods)**

Binomial linear mixed-effect model, fitting the interaction between the 3 factors: *news engagement*, *rank* and *representativeness*. We controlled for individual characteristics (age, sex, region, education, political alignment), an election which occurred during data collection, and we defined random intercepts for repeated measures (participant, search page, and news domain).

|  | is_selected_result | | |
| --- | --- | --- | --- |
| Predictors | Odds Ratios | CI | p |
| (Intercept) | 0.23 | 0.21 – 0.25 | <0.001 |
| engagement TD C | 0.99 | 0.91 – 1.08 | 0.844 |
| inv rank | 3.09 | 2.83 – 3.38 | <0.001 |
| news representativeness | 0.74 | 0.65 – 0.85 | <0.001 |
| is female | 0.99 | 0.92 – 1.06 | 0.801 |
| is west germany | 0.95 | 0.88 – 1.02 | 0.143 |
| age | 1.01 | 0.93 – 1.09 | 0.810 |
| education | 0.95 | 0.88 – 1.02 | 0.163 |
| political alignment | 1.04 | 0.97 – 1.12 | 0.284 |
| is during elections | 0.95 | 0.88 – 1.02 | 0.132 |
| is after elections | 1.00 | 0.94 – 1.08 | 0.913 |
| engagement TD C * inv rank | 0.98 | 0.90 – 1.07 | 0.669 |
| engagement TD C * news representativeness | 1.04 | 0.95 – 1.14 | 0.399 |
| inv rank * news representativeness | 1.15 | 1.00 – 1.31 | **0.047** |
| (engagement TD C * inv rank) * news representativeness | 0.98 | 0.88 – 1.08 | 0.624 |
| **Random Effects** | | | |
| $\sigma^2$ | 3.29 | | |
| $\tau_{00}$ search_page | 0.00 | | |
| $\tau_{00}$ participant | 0.00 | | |
| $\tau_{00}$ coredomain | 0.01 | | |
| N search_page | 1205 | | |
| N coredomain | 178 | | |
| N participant | 275 | | |
| Observations | 5323 | | |
| Marginal $R^2$ / Conditional $R^2$ | 0.281 / NA | | |

**S6. Full regression model for engagement calculated on the Google independant dataset (see D dataset in the Data Sources section of the of the Methods)**

Binomial linear mixed-effect model, fitting the interaction between the 3 factors: *news engagement*, *rank* and *representativeness*. We controlled for individual characteristics (age, sex, region, education, political alignment), an election which occurred during data collection, and we defined random intercepts for repeated measures (participant, search page, and news domain).

|  | is_selected_result | | |
| --- | --- | --- | --- |
| Predictors | Odds Ratios | CI | p |
| (Intercept) | 0.23 | 0.21 – 0.26 | **<0.001** |
| engagement TD D | 1.07 | 0.99 – 1.16 | 0.076 |
| inv rank | 3.09 | 2.83 – 3.39 | **<0.001** |
| news representativeness | 0.74 | 0.65 – 0.85 | **<0.001** |
| is female | 1.00 | 0.93 – 1.07 | 0.914 |
| is west germany | 0.95 | 0.88 – 1.02 | 0.159 |
| age | 1.01 | 0.93 – 1.09 | 0.857 |
| education | 0.95 | 0.88 – 1.02 | 0.145 |
| political alignment | 1.05 | 0.97 – 1.12 | 0.223 |
| is during elections | 0.95 | 0.88 – 1.02 | 0.136 |
| is after elections | 1.00 | 0.93 – 1.07 | 0.968 |
| engagement TD D * inv rank | 0.98 | 0.90 – 1.06 | 0.587 |
| engagement TD D * news representativeness | 0.98 | 0.90 – 1.07 | 0.660 |
| inv rank * news representativeness | 1.15 | 1.01 – 1.31 | **0.040** |
| (engagement TD D * inv rank) * news representativeness | 0.92 | 0.83 – 1.02 | 0.114 |
| **Random Effects** | | | |
| $\sigma^2$ | | 3.29 | |
| $\tau_{00}$ search_page | | 0.00 | |
| $\tau_{00}$ participant | | 0.00 | |
| $\tau_{00}$ coredomain | | 0.01 | |
| N search_page | | 1205 | |
| N coredomain | | 178 | |
| N participant | | 275 | |
| Observations | | 5323 | |
| Marginal $R^2$ / Conditional $R^2$ | | 0.283 / NA | |